\documentclass[reprint,english,aps,prl,twocolumn,amsmath, amssymb, showpacs,showkeys]{revtex4-2}
\usepackage[T1]{fontenc}
\usepackage{subfigure}
\usepackage[latin1]{inputenc}
\usepackage{graphicx}
\usepackage{epstopdf}
\usepackage{epsfig}
\usepackage{prettyref}
\usepackage{babel}
\usepackage{color}
\usepackage{amsmath}
\usepackage{float}
\usepackage[colorlinks=true,allcolors=blue]{hyperref}
\usepackage[nameinlink,noabbrev,capitalize]{cleveref}
\usepackage{tabularx}
\usepackage[table]{xcolor}
\usepackage{colortbl}

\makeatother

\begin{document}
\title{How to enhance anomalous Hall effects in magnetic Weyl semimetal Co$_3$Sn$_2$S$_2$ ?}

\author{Shivam Rathod}
\affiliation{UGC-DAE Consortium for Scientific Research, University Campus, Khandwa Road, Indore-452001, India}

\author{Megha Malasi}
\affiliation{UGC-DAE Consortium for Scientific Research, University Campus, Khandwa Road, Indore-452001, India}

\author{Archana Lakhani}
\affiliation{UGC-DAE Consortium for Scientific Research, University Campus, Khandwa Road, Indore-452001, India}

\author{Devendra Kumar}
\email{deveniit@gmail.com}\affiliation{UGC-DAE Consortium for Scientific Research, University Campus, Khandwa Road, Indore-452001, India}

\begin{abstract}

Large spin-orbit coupling, kagome lattice, nontrivial topological band structure with inverted bands anti-crossings, and Weyl nodes are essential ingredients, ideally required to obtain maximal anomalous Hall effect (AHE) are simultaneously present in Co$_3$Sn$_2$S$_2$. It is a leading platform to show large intrinsic anomalous Hall conductivity (AHC) and giant anomalous Hall angle (AHA) simultaneously at low fields. The giant  AHE in Co$_3$Sn$_2$S$_2$ is robust against small-scale doping-related chemical potential changes. In this work, we unveil a selective and co-chemical doping route to maximize AHEs in Co$_3$Sn$_2$S$_2$.  To begin with, in  Co$_3$Sn$_{2-x}$In$_x$S$_2$, we brought the chemical potential at the hotspot of Berry curvature along with a maximum of asymmetric impurity scattering in high mobility region. As a result at x=0.05, we found a significant enhancement of AHA (95\%) and AHC (190\%) from the synergistic enhancement of extrinsic and intrinsic mechanisms from modified Berry curvature of gaped nodal lines. Later, with anticipation of further improvements in AHE, we grew hole-co-doped Co$_{3-y}$Fe$_y$Sn$_{2-x}$In$_x$S$_2$ crystals, where we found rather a  suppression of AHEs. The role of dopants in giving extrinsic effects or band broadening can be better understood when chemical potential does not change after doping. By simultaneous and equal co-doping with electrons and holes in Co$_{3-y-z}$Fe$_y$Ni$_z$Sn$_2$S$_2$, we kept the chemical potential unchanged.  Henceforth, we found a significant enhancement in intrinsic AHC $\sim$116\% due to the disorder broadenings in kagome bands.

\end{abstract}

\maketitle
\section{Introduction}
Co$_3$Sn$_2$S$_2$ is a recently discovered magnetic Weyl semimetal with a Curie temperature (T$_C$) $\sim$177K, which is gaining attention due to its giant AHE~\cite{1Liu2018}. Conventionally~\cite{2Nagaosa2010}, ferromagnets with strong magnetization are expected to show large Hall resistivity given by $\rho_{xy}$=$\rho^0_{xy}$+$\rho^A_{xy}$ = R$_0$B+4$\pi$R$_s$M, where $\rho^0_{xy}$ and $\rho^A_{xy}$ are ordinary and anomalous Hall resistivity which depend on magnetic field (B) and magnetization (M) respectively, R$_0$ and R$_s$ are ordinary and anomalous Hall coefficients~\cite{3Jena2020}. Although Co$_3$Sn$_2$S$_2$ exhibits weak magnetization compared with typical magnetic systems, still it dominantly leads~\cite{4Shen2020} AHEs with large anomalous Hall factor (S$_H$)$\approx$1.1V$^{-1}$, defined as AHC ($\sigma^A_{xy}$) per unit moment (S$_H$=$\sigma^A_{xy}$/M).

In ferromagnets, intrinsic mechanisms~\cite{5Karplus1954} related to the Berry curvature of the bands and extrinsic mechanisms (side jump~\cite{6Berger1970} and skew scattering~\cite{7Smit1955}) involving spin-orbit interaction of magnetic electrons are generally considered to be the origin of AHE~\cite{2Nagaosa2010}. The intrinsic contribution to AHC ($\sigma^A_{xy,in}$) from Berry curvature ($\Omega^z_n$(\textbf{k})) of occupied band n is given by $\sigma^A_{xy,in}$= -e$^2$/$\hbar$$\int_{BZ}$d\textbf{k}/(2$\pi$)$^3$$\Sigma_n$\textit f$_n($\textbf{k}$)\Omega^z_n$(\textbf{k}),  where \textit f$_n($\textbf{k}) is the Fermi-Dirac distribution function~\cite{8Xiao2010}. Co$_3$Sn$_2$S$_2$ exhibits a large intrinsic AHC ($\sigma^A_{xy,in}$) $\sim$500-1400$\Omega^{-1}$cm$^{-1}$ as well as a giant AHA ($\sigma^A_{xy}$/$\sigma_{xx}$)$\times$100 $\sim$11-32\%, making it the leader of the known AHE materials~\cite{1Liu2018,9Tanaka2020,10Wang2018,11Li2020}. In magnetic Weyl semimetals, because of large $\sigma^A_{xy,in}$ and low longitudinal conductivity ($\sigma_{xx}$) from vanishing carrier density near the Weyl nodes, a large AHA is expected~\cite{1Liu2018}.

Co$_3$Sn$_2$S$_2$ is a weak itinerant half metallic ferromagnet~\cite{26Weihrich2006,27Solovyev2021}, when doped with Ni~\cite{28Kubodera2006} or Fe~\cite{29Kassem2013} for Co, Se~\cite{30Sakai2013} for S, and In~\cite{26Weihrich2006} for Sn, T$_C$ and magnetization gradually decrease. Application of pressure~\cite{31Chen2019,32Zeng2022,33Liu2020,34Guguchia2020} reduces magnetization~\cite{25Armitage2018,35Burkov2014} which decreases the separation of Weyl nodes (k$_W$) and associated AHC ($\sigma^A_{xy,in}$= k$_W$e$^2$/$\pi$h). In Co$_3$Sn$_2$S$_2$, enhanced AHC can be obtained by  topological Hall effect~\cite{36Neubauer2022} from non-collinear magnetic structures or coexisting ferromagnetic and helimagnetic phases~\cite{37Guguchia2021}. AHE is suppressed~\cite{38Ikeda2021} in thinner films ($\textless$20nm) of lower quality~\cite{11Li2020,39Shiogai2021} while enhances in nanoflakes~\cite{40Yang2020} with high mobility~\cite{9Tanaka2020}.

\begin{figure*}[!t]
\begin{centering}
\includegraphics[scale=1]{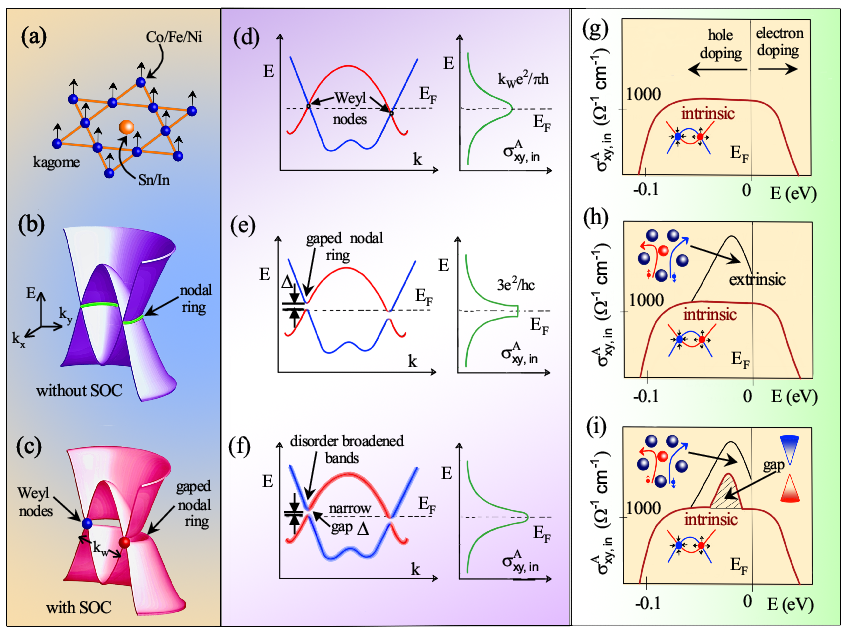}
\par\end{centering}
\caption{(Colour Online) (a) In Co$_3$Sn$_2$S$_2$, magnetic Co atoms form a kagome lattice structure whose center is Sn; dopants Fe and Ni substitute Co while In substitutes at Sn sites. (b) Schematic piece of the band structure of Co$_3$Sn$_2$S$_2$ near E$_F$ forming a nodal ring which is converted to (c) gapped nodal rings (anticrossings) everywhere on the ring except at Weyl node crossings~\cite{1Liu2018} upon including the effect of spin-orbit coupling (SOC). Intrinsic AHC ($\sigma^A_{xy,in}$) is related to hot spots of bands Berry curvature dominantly coming from (d) Weyl nodes monopoles which depend on the separation of Weyl nodes (k$_W$)  as $\sigma^A_{xy,in}$= k$_W$e$^2$/$\pi$h, (e) gapped ($\Delta$) nodal ring at E$_F$ giving a quantized $\sigma^A_{xy,in}$=3e$^2$/hc from massive Dirac gap~\cite{21Zhou2020}, and (f) broadening (shown by spread) of bands including nodal rings at E$_F$ due to doping disorder, producing many overlapping~\cite{19Markou2021,22Bianco2014} which further increases the $\sigma^A_{xy,in}$ due to a narrow~\cite{23Shen2020}  gap ($\Delta$). (g) Co$_3$Sn$_2$S$_2$ exhibit~\cite{1Liu2018,23Shen2020} plateau (red curve) of large $\sigma^A_{xy,in}$$\textgreater$1000$\Omega^{-1}$cm$^{-1}$ in a wide region ($\approx$100meV) around E$_F$, including contributions from Weyl nodes. (h)  The addition of Fe~\cite{23Shen2020} or In~\cite{21Zhou2020} causes hole doping and shifts the chemical potential (\textmu) below the pristine case. The doping causes extrinsic impurity scattering which gives a mountain (black curve) of increased AHC. (i)  For In doping, due to modified~\cite{24Yanagi2021} Berry curvature of gapped nodal rings/lines and massive Dirac gap~\cite{21Zhou2020}, $\sigma^A_{xy,in}$ gets enhanced (shaded red mountain) along with enhanced extrinsic impurity scattering. The vertical dotted line in Fig (g-i) represents the Fermi energy of pristine Co$_3$Sn$_2$S$_2$. } \label{Fig1}
\end{figure*}

In Co$_3$Sn$_2$S$_2$ bands at E$_F$ mainly originate~\cite{12Li2019} from the Co layers forming a kagome structure as shown in Fig.~\ref{Fig1}(a). A kagome structure naturally gives rise to band crossing~\cite{13Wang2013,14Ye2018} and flat bands~\cite{15Jiang2021}. Further strong SOC and electronic correlations~\cite{16Yazyev2019,17Liu2019,18Yin2019} cause multiple band overlaps~\cite{19Markou2021} which enhance Berry curvature~\cite{20Asaba2021}. AHE in Co$_3$Sn$_2$S$_2$ is governed by large Berry curvature from band structure~\cite{1Liu2018} at E$_F$ related to Co kagome as shown by the schematic in Fig.~\ref{Fig1}(b). In the absence~\cite{10Wang2018} of SOC, valence and conduction bands inverse around the L point and form a nodal ring protected by mirror symmetry M(010), and when SOC is included this nodal ring becomes gapped as shown in Fig.~\ref{Fig1}(c) with a pair of Weyl nodes $\sim$60 meV above~\cite{1Liu2018} E$_F$. These Weyl points act like magnetic monopoles in momentum space and serve as a large source of Berry Curvature~\cite{25Armitage2018}. As shown in Fig.~\ref{Fig1}(d-e), due to the presence of Weyl nodes and gapped nodal rings near E$_F$, a plateau [Fig.~\ref{Fig1}(g)] of large intrinsic AHC with $\sigma^A_{xy,in}$\textgreater1000$\Omega^{-1}$cm$^{-1}$ in a broad region ($\sim$100meV) around E$_F$ is predicted~\cite{1Liu2018}, which is robust against small-scale thermal and chemical potential changes~\cite{23Shen2020}. Assuming a rigid band, this suggests that the intrinsic AHC in Co$_3$Sn$_2$S$_2$ is already at its maximum and cannot be further enhanced by electron or hole doping~\cite{4Shen2020}. However, this gives us an ideal doping platform to enhance these already giant effects through modulation of nodal line gap and extrinsic effects.

Recent studies~\cite{4Shen2020,21Zhou2020,23Shen2020,41Thakur2020,42McGuire2021} of doping have shown that the AHE in Co$_3$Sn$_2$S$_2$ is enhanced due to the appearance of extrinsic mechanisms from impurity scattering. The extrinsic mechanisms are due to a sudden change in periodic potential experienced by electrons due to dopants or lattice defects~\cite{4Shen2020}. In comparison to indium~\cite{21Zhou2020} at the tin site ($\sim$800$\Omega^{-1}$cm$^{-1}$) or nickel doping~\cite{4Shen2020} at conducting cobalt kagome lattice site ($\sim$214$\Omega^{-1}$cm$^{-1}$), iron doping~\cite{23Shen2020} significantly enhances extrinsic AHC by $\sim$1071$\Omega^{-1}$cm$^{-1}$ and also AHA to a maximum of 33\% (from giant skewness). Further extrinsic effects get enhanced with mobility~\cite{9Tanaka2020} and decrease at higher doping~\cite{21Zhou2020} as shown by a black mountain in Fig.~\ref{Fig1}(h).

Recent experimental~\cite{21Zhou2020} and theoretical~\cite{24Yanagi2021} studies suggest that the AHE in Co$_3$Sn$_2$S$_2$ can be maximized if we tune the chemical potential to a maximum of the modified~\cite{24Yanagi2021} Berry curvature of gaped nodal lines or massive Dirac gap~\cite{14Ye2018,21Zhou2020} shown with shaded mountain Fig.~\ref{Fig1}(i) using In dopings and simultaneously maximize asymmetric scattering from Fe~\cite{23Shen2020} impurities in the high mobility~\cite{9Tanaka2020,21Zhou2020} regime. To precisely place the chemical potential at the hotspot, first, we prepare hole~\cite{29Kassem2013} doped series Co$_3$Sn$_{2-x}$In$_x$S$_2$ to enhance the AHEs of pristine from its given value (plateau). In practice, the exact location of these hotspots is unpredictable~\cite{4Shen2020} and varies due to complicated growth-dependent crystal disorders/doping defects, better detected experimentally. Experimentally, at x=0.05 we found maximum enhancements in AHC ~190\%. Thereafter, to get further enhanced extrinsic effects from Fe~\cite{23Shen2020} and to simultaneously pin the E$_F$ at the maximal hotspot at the plateau of the intrinsic regime shown in Fig.~\ref{Fig1}(i), low hole-doped Co$_{3-y}$Fe$_y$Sn$_{2-x}$In$_x$S$_2$ at (0$\le$x=y$\le$0.05) crystals are prepared. However, we found rather a suppression of AHC.

Recent studies~\cite{4Shen2020} show that dopants increase disorder in the system and cause the broadening of bands as shown by a spread of bands at E$_F$ in Fig.~\ref{Fig1}(f). It reduces the nodal ring gap ($\Delta$) and the intrinsic effects get enhanced~\cite{4Shen2020,19Markou2021,22Bianco2014}. Theoretically~\cite{4Shen2020}, Co$_{3-z}$Ni$_z$Sn$_2$S$_2$ at z=0.056 would produce maximum enhancement due to disorder. In Co$_3$Sn$_2$S$_2$, the E$_F$ is already located at the maxima of Berry curvature and in recent studies, the position of chemical potential is shifted to higher or lower energy. However, no attempts are made to keep the chemical potential fixed by simultaneous and equal electron and hole co-doping. This will better explain the scenario of disorder-broadened bands and extrinsic scattering effects. Also, at higher nickel/electron doping, the chemical potential shifts from the hotspot~\cite{4Shen2020} or plateau~\cite{41Thakur2020}, and  AHC is reduced, so the simultaneous addition of iron would prevent this as well. To address this issue, we have grown Co$_{3-y-z}$Fe$_y$Ni$_z$Sn$_2$S$_2$ and at y=z=0.025, the chemical potential remains fixed at the plateau and we found a notably large 116\% enhancement due to disorder broadening.

\section{Experimental Details}

Single crystals of Co$_3$Sn$_{2-x}$In$_x$S$_2$, Co$_{3-y}$Fe$_y$Sn$_{2-x}$In$_x$S$_2$, and Co$_{3-y-z}$Fe$_y$Ni$_z$Sn$_2$S$_2$ are grown by the self-flux method using the modified Bridgeman technique~\cite{44Rathodd2020,45Rathod2020}. The high quality of grown single crystals is examined by x-ray diffraction (XRD), high-resolution x-ray diffraction (HRXRD), and energy dispersive x-ray spectroscopy (EDS). Standard four-probe AC Resistivity and five-probe Hall measurements are performed on the polished rectangular bar-shaped polished crystals in 9 Tesla PPMS from Quantum Design.
\section{Results and Discussion}

\begin{figure}[]
\begin{centering}
\includegraphics[scale=0.8]{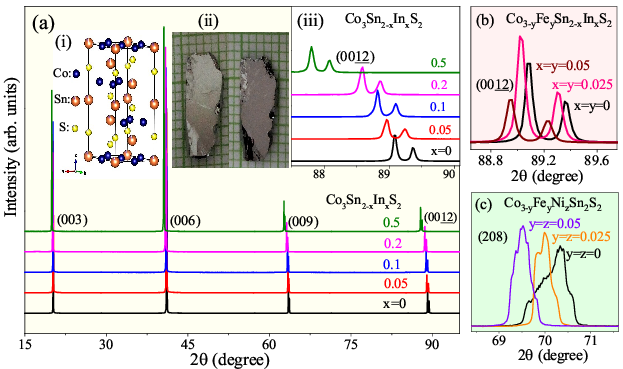}
\par\end{centering}
\caption{(Colour Online) (a) HRXRD $\theta$-2$\theta$ scan of Co$_3$Sn$_{2-x}$In$_x$S$_2$ single crystals indexed with (0 0 l). The crystal structure of Co$_3$Sn$_2$S$_2$ is shown in inset (i). Inset (ii) is the image of a cleaved Co$_3$Sn$_2$S$_2$ crystal. Inset (iii) is the magnified view of the (00\underline{12}) peak. (b) HRXRD $\theta$-2$\theta$ scan of (00\underline{12}) planes of Co$_{3-y}$Fe$_y$Sn$_{2-x}$In$_x$S$_2$ and (c) is the in-plane $\theta$-2$\theta$ scan of (208) crystal planes of Co$_{3-y-z}$Fe$_y$Ni$_z$Sn$_2$S$_2$.} \label{Fig2}
\end{figure}
Figure~\ref{Fig2}(a) shows the HRXRD pattern of Co$_3$Sn$_{2-x}$In$_x$S$_2$ single crystals at different x with peaks only from (0 0 l) planes showing that crystals are grown along the c axis. The inset (i) shows the crystal structure of Co$_3$Sn$_2$S$_2$ generated using VESTA software. Blue Co atoms form a kagome layer structure which leads to the conduction process and magnetism. Inset (ii) is the optical image of a pristine crystal cut into two with typical dimensions of 10$\times$5mm and a smooth shiny surface. Inset (iii) is the magnified view of the (00\underline{12}) peak and Fig.~\ref{Fig2}(b) for Co$_{3-y}$Fe$_y$Sn$_{2-x}$In$_x$S$_2$ shows a shift toward a lower angle in 2$\theta$ due to the expansion of unit cell along c with doping. Figure~\ref{Fig2}(c)  is the in-plane $\theta$-2$\theta$  scan of (208) crystal planes of Co$_{3-y-z}$Fe$_y$Ni$_z$Sn$_2$S$_2$. confirming the proper orientation of crystal planes along other crystallographic directions~\cite{45Rathod2020,46Kumar2015}. The systematic change of lattice parameters~\cite{21Zhou2020,29Kassem2013,47Sakai2015,48Kumar2017} and stoichiometric site occupancy observed from Rietveld refinement and atomic proportion in EDS spectra shows high-quality and single-phase of crystals. For more details see supplementary information.

\begin{figure}[]
\begin{centering}
\includegraphics[scale=0.51]{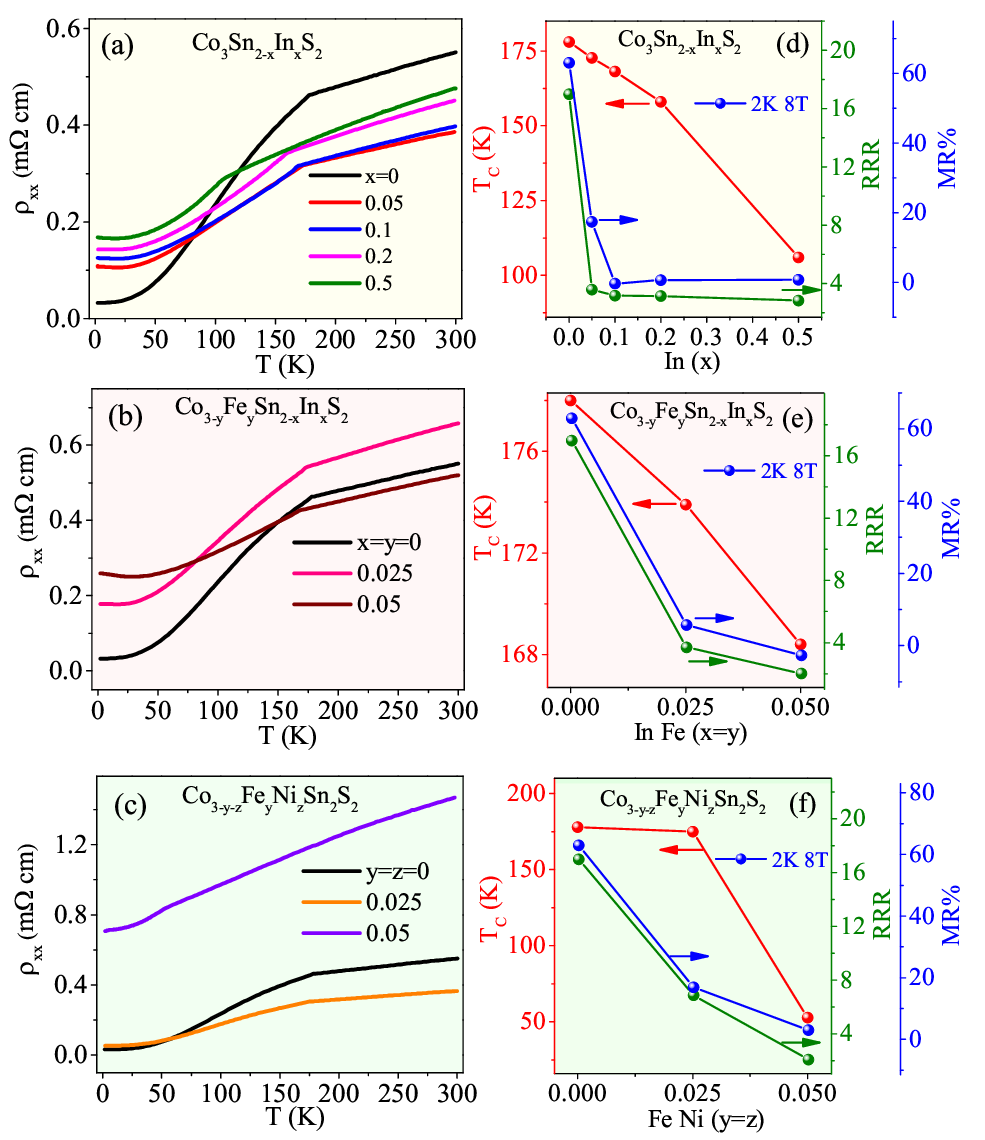}
\par\end{centering}
\caption{(Colour Online) Temperature dependence of resistivity ($\rho$$_{xx}$) of (a) Co$_3$Sn$_{2-x}$In$_x$S$_2$ for x =0 to 0.5, (b) Co$_{3-y}$Fe$_y$Sn$_{2-x}$In$_x$S$_2$ at x=y=0 to 0.05 (c) Co$_{3-y-z}$Fe$_y$Ni$_z$Sn$_2$S$_2$ at y=z=0 to 0.05. (d-f) Decrease of Curie temperature (T$_C$), residual resistivity ratio (RRR), and magnetoresistance\% (MR\%=\{[$\rho$$_{xx}$(8T)-$\rho$$_{xx}$(0T)]/ $\rho$$_{xx}$(0T)\}$\times$100) at 2K,  8Tesla with the increase in doping. } \label{Fig3}
\end{figure}

The temperature-dependent resistivity exhibits metallic behavior shown in Fig.~\ref{Fig3}(a-c). The co-doping results in a faster variation of residual resistivity $\rho$$_{xx}$(2K) compared to adding individual dopant impurities. The d$\rho$$_{xx}$/dT shows a sharp change in slope at T$_C$.  RRR=$\rho$$_{xx}$(300K)/$\rho$$_{xx}$(2K) is a measure of crystal defect density, follows MR\% as shown in ~\ref{Fig3}(d-f) and falls significantly with dopants impurities. For y=z=0.05, we observe a sudden drop in T$_C$ deviating from linearity, whereas x and x=y doping exhibit linear behavior.  Iron and nickel dopants weaken exchange couplings~\cite{47Sakai2015} and spin splittings~\cite{4Shen2020} by the filling of electrons in minority spin band~\cite{26Weihrich2006,38Ikeda2021}, it can go to exchange-split gap located at higher energy~\cite{29Kassem2013}. Changes in lattice parameters are consistent with previous reports~\cite{21Zhou2020,29Kassem2013,41Thakur2020,47Sakai2015}. Altogether, large RRR, T$_C$, MR\%, and low $\rho$$_{xx}$(2K) indicate~\cite{45Rathod2020,49Ding2019} higher crystal quality of Co$_3$Sn$_2$S$_2$ in comparison to reports~\cite{1Liu2018,10Wang2018,21Zhou2020,23Shen2020,41Thakur2020,50Shen2022}.  For more details see SI.
\begin{figure*}[]
\begin{centering}
\includegraphics[scale=0.5]{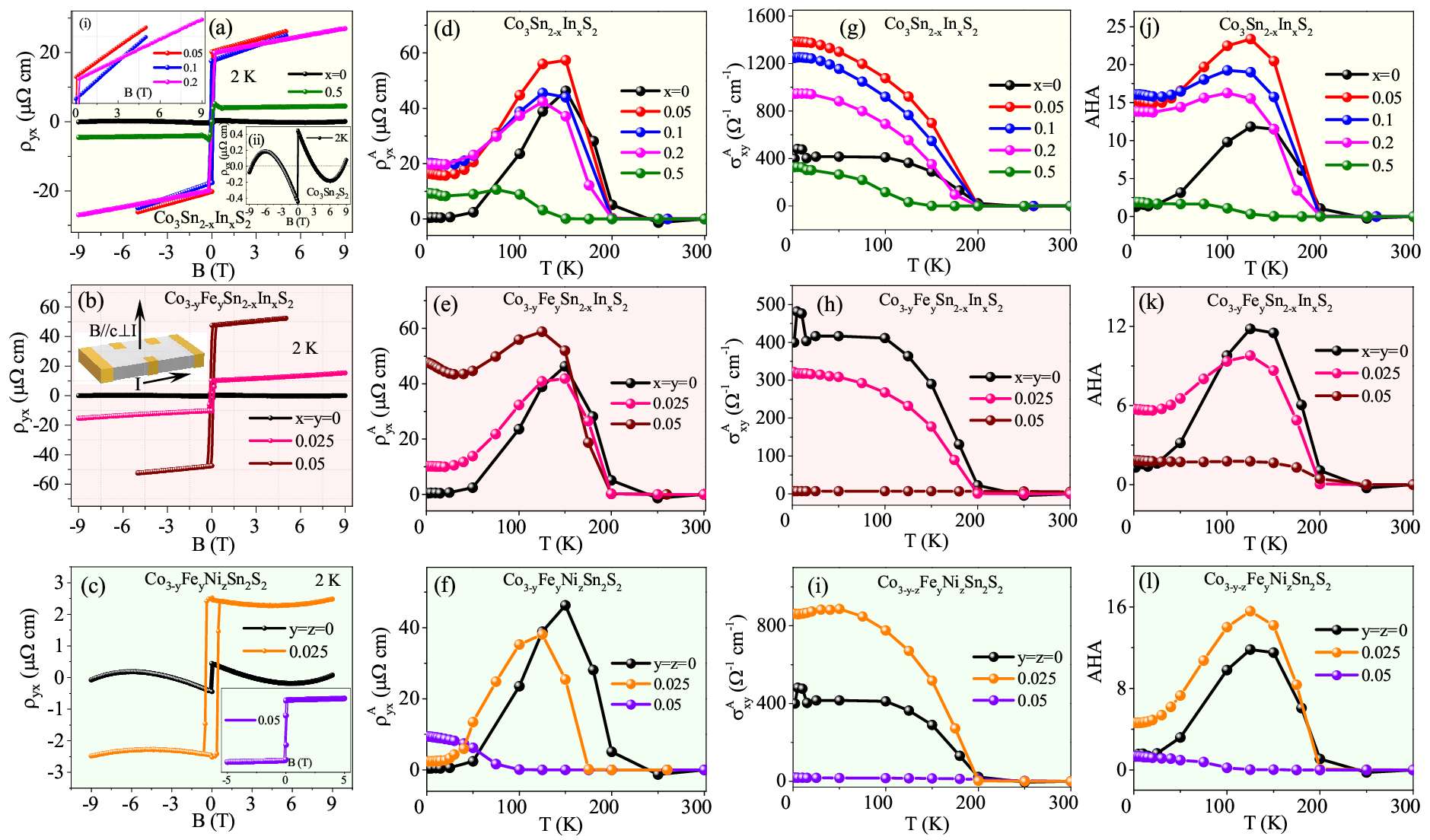}
\par\end{centering}
\caption{(Colour Online) Hall resistivity ($\rho$$_{yx}$) at 2 K for (a) Co$_3$Sn$_{2-x}$In$_x$S$_2$; inset (i) shows a magnified view of linearity for x=0.05, 0.1, and 0.2 while inset (ii) shows the nonlinearity for x=0, (b) Co$_{3-y}$Fe$_y$Sn$_{2-x}$In$_x$S$_2$,  inset shows the schematic of Hall contacts on the sample, the magnetic field (B) is applied along an easy c-axis perpendicular to current (I) (B//c$\perp$ I), and (c) Co$_{3-y-z}$Fe$_y$Ni$_z$Sn$_2$S$_2$ showing nonlinearity at y=z=0 and 0.025 whereas x=0.05 is linear.  Temperature dependence of (d-f) anomalous Hall resistivity ($\rho^A_{xy}$), (g-i) anomalous Hall conductivity ($\sigma^A_{xy}$), and (j-l) anomalous Hall angle AHA=($\sigma^A_{xy}$/$\sigma_{xx}$)$\times$100.} \label{Fig4}
\end{figure*}

At a very low saturating field with the saturation of magnetization, Hall resistivity saturates demonstrating an anomalous Hall effect in Fig.~\ref{Fig4}(a-c). The high-field Hall resistivity extrapolated to zero fields is called anomalous Hall resistivity ($\rho^A_{xy}$), shown in Fig.~\ref{Fig4}(d-f). Around T$_C$, the $\rho^A_{xy}$ exhibits a peak, which shifts toward low temperatures with an increase in doping. The non-zero values of  $\rho^A_{xy}$ for T\textgreater T$_C$ are due to field-enhanced paramagnetic short-range order~\cite{51Pickel2010,52Antropov2005}. The temperature-dependent large AHC $\sigma^A_{xy}$= $\rho$$_{yx}$/[$\rho$$_{yx}$$^2$ + $\rho$$_{xx}$$^2$] shown in Fig.~\ref{Fig4}(g-i) is a manifestation of the topological band structure.

For pristine x=0 and y=z=0.025 samples, large AHC, independent of temperature in the range of 2K to 100K suggests their origin is dominated by the intrinsic Berry curvature~\cite{2Nagaosa2010} which does not change significantly with temperature because chemical potential does not change appreciably for small thermal energy. At higher temperatures (T$\rightarrow$T$_C$) reduction in AHC is due to a decrease in saturation magnetization and Berry curvature~\cite{10Wang2018,34Guguchia2020}. In comparison to the pristine (x=0) sample, the large AHC at x=0.05, 0.1, 0.2, and y=z=0.025 suggest chemical potential (\textmu) is at the plateau, while it falls off at higher dopings.

The AHA measures how much current flows in the y direction when we apply current in the x direction, it, therefore, measures the efficiency of a Hall device. The magnetic Weyl semimetals due to Weyl nodes dominated large $\sigma^A_{xy}$ and simultaneously due to low carrier density induced small $\sigma_{xx}$ are expected~\cite{1Liu2018} to show large AHA. The intrinsic $\sigma^A_{xy,in}$ is robust and temperature independent but the $\sigma_{xx}$ decreases with increasing temperature due to contributions from scattering mechanisms~\cite{3Jena2020,53Rathod2022}. Therefore, for intrinsic dominated $\sigma^A_{xy}$, the AHA will show a peak below T$_C$ which is observed for x=0 ($\approx$12\%) and y=z=0.025 ($\approx$16\%) shown in Fig.~\ref{Fig4}(j-l). At low temperatures, the AHA is improved for low doping (x=0.05, 0.1, 0.2, y=z=0.025) and decreases at higher doping.

\begin{figure*}[]
\begin{centering}
\includegraphics[scale=0.7]{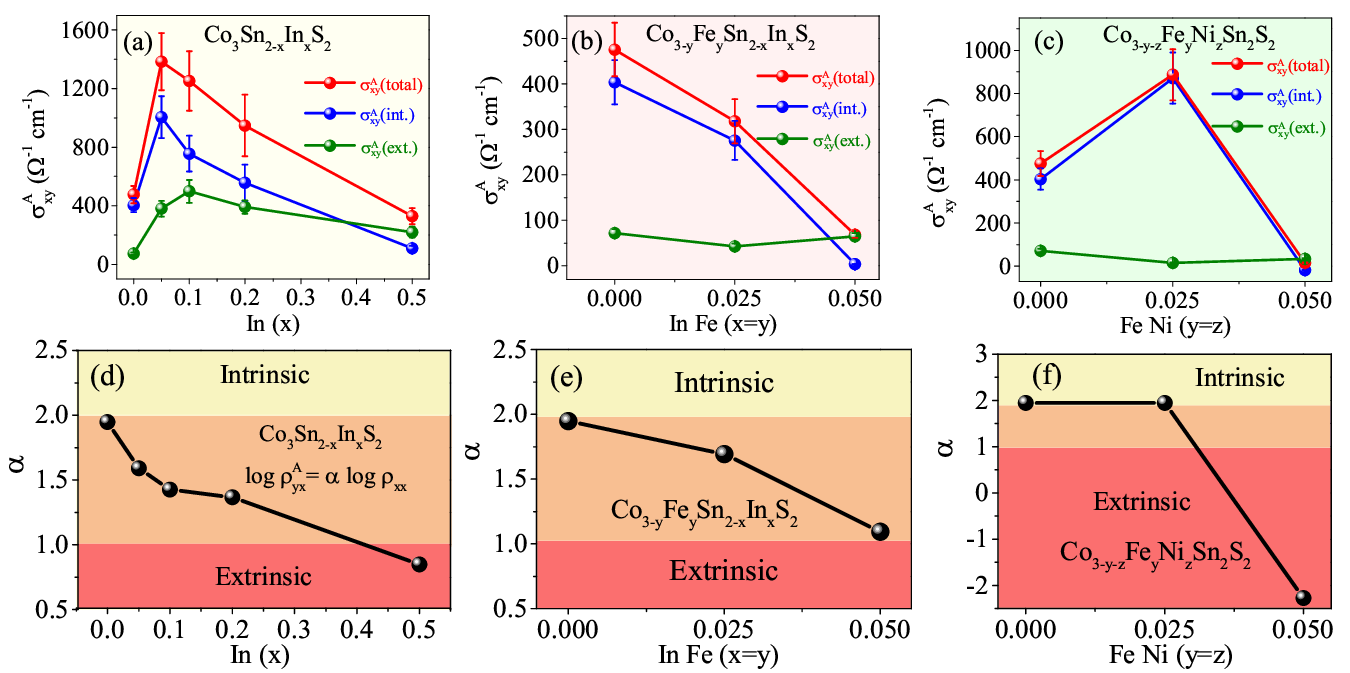}
\par\end{centering}
\caption{(Colour Online) The plot of total anomalous Hall conductivity ($\sigma^A_{xy}$) shown with red balls, intrinsic anomalous Hall conductivity($\sigma^A_{xy,in}$ ) blue balls, and extrinsic anomalous Hall conductivity ($\sigma^A_{xy,ex}$) green balls for (a) Co$_3$Sn$_{2-x}$In$_x$S$_2$, (b) Co$_{3-y}$Fe$_y$Sn$_{2-x}$In$_x$S$_2$, and (c) Co$_{3-y-z}$Fe$_y$Ni$_z$Sn$_2$S$_2$ obtained from TYJ scaling~\cite{54Tian2009}. The error bars are due to the sample dimensions measurement uncertainty (11-22\%) (d) The plot of alpha ($\alpha$) vs x, (e) alpha ($\alpha$) vs x,y and (f) alpha ($\alpha$) vs y,z obtained through a linear fit of log$\rho^A_{yx}$ vs log$\rho_{xx}$; The yellow region ($\alpha$$\ge$2)  represents the dominant intrinsic mechanism, red region ($\alpha$$\le$1) represents the Hall resistivity dominated by the extrinsic mechanism and light red (1\textless$\alpha$\textless2) represent the contribution from both.[see SI]} \label{Fig5}
\end{figure*}

The intrinsic and extrinsic mechanisms shown in Fig.~\ref{Fig5}(a-c) are separated by Tian-Ye-Jin (TYJ) scaling~\cite{54Tian2009} model. In the TYJ model, the conductivity equation is $\sigma^A_{xy}$=-a$\sigma^{-1}_{xx0}$$\sigma^2_{xx}$-b, where a represents extrinsic, b represents intrinsic contributions, and $\sigma_{xx0}$ is the residual conductivity. For the pristine x=0 sample, intrinsic AHC predominates over extrinsic $\sigma^A_{xy,ex}$=72$\pm$9$\Omega^{-1}$cm$^{-1}$ but the obtained value of $\sigma^A_{xy,in}$= 404$\pm$49$\Omega^{-1}$cm$^{-1}$ is much lower than expected~\cite{1Liu2018}. A similar low value of $\sigma^A_{xy,in}$=505$\Omega^{-1}$cm$^{-1}$ is previously observed~\cite{10Wang2018}. For crystals grown using a particular method, $\sigma^A_{xy}$ does not vary much for a relative angle of field and c-axis, thickness, or magnetization~\cite{10Wang2018}. We have performed measurements on another high-quality Co$_3$Sn$_2$S$_2$ crystal with similar RRR giving $\sigma^A_{xy}$=524$\pm$64$\Omega^{-1}$cm$^{-1}$. Depending on crystal quality and growth methods a large variation in residual resistivity~\cite{1Liu2018,10Wang2018,49Ding2019} $\rho_{xx}$(2K) and $\sigma^A_{xy}$ can be seen in reports~\cite{1Liu2018,9Tanaka2020,10Wang2018,11Li2020,schilberth2023nodal}.

In Co$_3$Sn$_2$S$_2$, AHC from extrinsic side jump $\sigma^A_{xy,sj}=$[e$^2$/(ha)]$\times$(E$_{SO}$/E$_F$)=3.9$\Omega^{-1}$cm$^{-1}$ is negligible, where  E$_{SO}$ is spin-orbit interaction energy, a is the lattice constant~\cite{10Wang2018} . Therefore extrinsic AHC is mainly from skew scattering which is large in high mobility samples~\cite{55Onoda2006,56Onoda2008} $\sigma^A_{xy,sk}$=E$_{SO}$U$_{imp}$ $\sigma_{xx}$/W$^2$, where W is bandwidth, AHA increases with an increase in the strength of impurity potential~\cite{6Berger1970} U$_{imp}$. In Co$_3$Sn$_2$S$_2$, Weyl nodes separated by~\cite{10Wang2018} k$_W$=0.3\textup {\AA}$^{-1}$ at M=0.33\textmu$_B$/Co are expected~\cite{25Armitage2018,35Burkov2014} to give $\sigma^A_{xy,in}$=[e$^2$/(4$\pi^2$)]$\times$k$_W$=525$\Omega^{-1}$cm$^{-1}$.

As shown for Co$_3$Sn$_{2-x}$In$_x$S$_2$ in Fig.~\ref{Fig5}(a) at x=0.05, $\sigma^A_{xy,in}$ is significantly increased to a maximum of 1004$\Omega^{-1}$cm$^{-1}$ and further decreases to 755$\Omega^{-1}$cm$^{-1}$ at x=0.1, 557$\Omega^{-1}$cm$^{-1}$ at x=0.2, and 108$\Omega^{-1}$cm$^{-1}$ at x=0.5. In comparison to x=0, at x=0.05, such a large drastic enhancement of intrinsic AHC by 148\% cannot be ascribed solely to the theoretically~\cite{24Yanagi2021} expected 11\% from modified Berry curvature around nodal lines. This suggests that for chemical potential at x=0.05, the nodal lines are also getting modified into a massive Dirac gap~\cite{14Ye2018} which gives a quantized value of $\sigma^A_{xy,in}$=3e$^2$/(hc)=881$\Omega^{-1}$cm$^{-1}$ for c=13.176\textup {\AA} and the prefactor 3 is from kagome layers per unit cell~\cite{21Zhou2020}. In addition, we observe that the mobility is large and the carriers are minimum near x=0.05, 0.1. With increasing x, the mobility and MR\% decrease while charge carriers increase indicating a shift of the chemical potential towards lower energy from nontrivial topological nodal line gaped region.[see SI] A large skew scattering of $\sigma^A_{xy,sk}$=380$\pm$54$\Omega^{-1}$cm$^{-1}$ observed at x=0.05 can be expected from a highly conductive semimetal due to geometric frustrations in the kagome lattice having Dirac quasiparticles~\cite{43Yang2020}. At the highest x=0.5 doping, the system went below the plateau.

As shown in Fig.~\ref{Fig5}(b) for Co$_{3-y}$Fe$_y$Sn$_{2-x}$In$_x$S$_2$, $\sigma^A_{xy,in}$ is reduced at x=y=0.025 and is completely suppressed at x=y=0.05. This is contrary to expectations since Fe and In dopings have the same effect on chemical potential~\cite{29Kassem2013}.  We observe that $\sigma^A_{xy,in}$ mimics carrier density. [see SI] This suggests that chemical potential is much more sensitive for Fe and In hole co-dopings and is leaving the plateau/hotspot and possibly co-dopings are cancelling~\cite{19Markou2021} the nodal line Berry curvature~\cite{schilberth2023nodal}.

As shown in Fig.~\ref{Fig5}(c) for Co$_{3-y-z}$Fe$_y$Ni$_z$Sn$_2$S$_2$ at y=z=0.025, the intrinsic AHC $\sigma^A_{xy,in}$ is notably enhanced by 116\% in comparison to the pristine sample. Equal electron (Ni) and holes (Fe) co-doping should keep \textmu~intact. As shown in inset (ii) of Fig.~\ref{Fig4}(a) for pristine x=0, the $\rho^A_{xy}$ shows highly nonlinear behavior due to co-existing electron and hole bands. This nonlinearity is extremely sensitive to temperature and disappears at T\textgreater50K and even for light chemical dopings related chemical potential changes as can be seen in inset(i) of Fig.~\ref{Fig4}(a-c). The observation of large nonlinearity in $\rho^A_{xy}$ for y=z=0.025 similar to the pristine sample indicates that we are successful in keeping the chemical potential intact.  Further, from two band fittings, we observe high mobility and low carrier semimetallic characteristics similar to the pristine sample[see SI]. Theoretically~\cite{4Shen2020} Co$_{3-z}$Ni$_z$Sn$_2$S$_2$ at z=0.056 would produce maximum enhancement due to disorder broadening. It is equivalent to our y=z=0.025 sample which shows a notably large 116\% enhancement in comparison to the 60\% reported~\cite{4Shen2020} for Ni. Fe and Ni dopants perhaps modify SOC~\cite{12Li2019}, local lattice potential/spacings, translational symmetry, and interband coupling~\cite{22Bianco2014}, leading to the broadening of kagome bands. This can result in a strong overlap between SOC-gaped nodal lines present in Co$_3$Sn$_2$S$_2$ which reduces the gap and gives harmonic contributions to enhance $\sigma^A_{xy,in}$. At higher dopings y=z =0.05 a negative~\cite{57Zhu2016} contribution to AHC is observed which is previously~\cite{23Shen2020} attributed to Kondo scattering due to iron.

 Furthermore, the intrinsic and extrinsic contributions determined by the TYJ model are in excellent agreement with those determined by the equation~\cite{58Mitra2010} log$\rho^A_{yx}$ vs log$\rho_{xx}$ as shown in Fig.~\ref{Fig5}(d-f) and also with Unified model~\cite{2Nagaosa2010,55Onoda2006,56Onoda2008,59Miyasato2007} shown in SI Fig. S8.

\begin{figure*}[]
\begin{centering}
\includegraphics[scale=0.65]{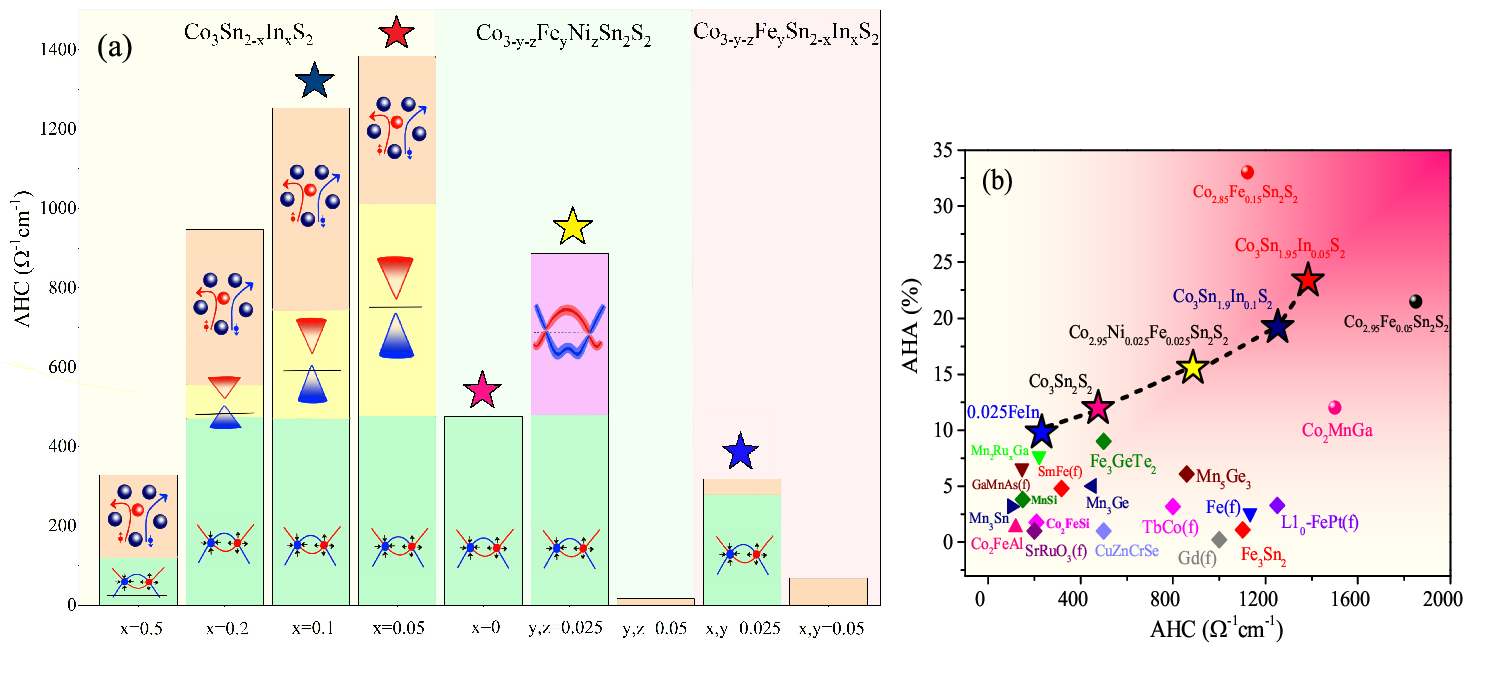}
\par\end{centering}
\caption{(Colour Online) (a) AHC of doped samples In(x), InFe(x=y), and FeNi(y=z) compared with pristine (x=0).  The green levels show the intrinsic contributions from Weyl nodes (chemical potential in the plateau region of Fig.~\ref{Fig1}(g)), the light red levels show the addition of extrinsic contribution on In doping (x\textgreater0), the yellow levels show the additional intrinsic contribution from modified gaped nodal lines (shaded red mountain regime in Fig.~\ref{Fig1}(i)). At higher In dopings \textmu leaves the hotspot/plateau. At y=z=0.025 the disorder-broadened kagome bands (shown with magenta) enhance AHC whereas chemical potential leaves the plateau for x=y=0.025 or 0.05. (b) Comparison of large AHA and AHC values obtained from this work shown by the trail of stars with other AHE systems. All these materials show large AHE at low fields suitable for applications~\cite{23Shen2020}. } \label{Fig6}
\end{figure*}

The large intrinsic AHC and AHA in Co$_3$Sn$_2$S$_2$ is a manifestation of its topological band structure making it lead among most materials.  As long as the E$_F$ is close to topological nodal lines crossing and anticrossing region, the AHC is large and exhibits a plateau as shown by green levels in Fig.~\ref{Fig6}(a). Further, these nodal lines regions can be modified even by small indium dopings and by reducing the SOC split anticrossings into a massive Dirac gap leading to significant enhancement in Berry curvature giving enhanced AHC shown by yellow levels in Fig.~\ref{Fig6}(a). On the other hand, Fe and Ni directly substitute the Co which is the building block of kagome, at light doping y=z=0.025 the AHC significantly elevated ~116\% due to disorder broadening of kagome bands, shown by violet levels in Fig.~\ref{Fig6}(a). For co-dopings, x=y or y=z the rapid change in AHC and chemical potential suggest that the system is much more responsive for co-dopings and the assumption of a rigid band is not justified for the broad doping range. The extrinsic scattering due to dopants is found to be large for indium, in comparison to co-dopings. Further from this work, we obtained large AHC and AHA values for Co$_3$Sn$_{1.95}$In$_{0.05}$S$_2$, Co$_3$Sn$_{1.9}$In$_{0.1}$S$_2$, Co$_{2.95}$Ni$_{0.025}$Fe$_{0025}$Sn$_2$S$_2$, Co$_3$Sn$_2$S$_2$, and  Co$_{2.975}$Fe$_{0.025}$Sn$_{1.975}$In$_{0.025}$S$_2$ doped samples shown by the trail of stars in Fig.~\ref{Fig6}(b). The leader doped-Co$_3$Sn$_2$S$_2$ group tending to the top right~\cite{23Shen2020} defeats other known materials~\cite{14Ye2018,59Miyasato2007,60Imort2012,61Nakatsuji2015,62Manyala2004,63Pu2008,64Belopolski2019,65Ruediger2000,66Kim2000,67Zeng2006,68Kim2018,69Kim2001,70Wu2010,71Fang2003} shown in the bottom left. The AHA represents the efficiency of getting transverse current with simultaneously large AHC values at zero fields making these materials potential candidates for applications in Hall sensors~\cite{72Zhang2019} or spin-orbit torques~\cite{73Iihama2018}. Altogether, this work demonstrates how selective Fe and Ni doping inside the kagome, In doping outside or co-doping, should be utilized to enhance AHE in Co$_3$Sn$_2$S$_2$.

\section{Conclusion}

We unveil a selective and co-doping route to enhance AHEs in Co$_3$Sn$_2$S$_2$.  To start with, by indium doping we brought the chemical potential in the region of modified Berry curvature and massive Dirac gap, found near x=0.05 giving enhanced AHC $\sim$190\% due to combined intrinsic and extrinsic effects. Further, when we brought chemical potential (\textmu) in the same region using x=y=0.025, the intrinsic effects rather decreased, suggesting extreme sensitivity of nodal lines and \textmu~for co-doping. Further in Co$_3$Sn$_2$S$_2$, we have shown that by co-doping with equal electrons and holes the chemical potential can be held intact. At y=z=0.025, we found a noteworthy enhancement of intrinsic AHC $\sim$116\% due to enhanced Berry curvature of SOC gaped nodal lines at E$_F$ from disorder broadening of bands caused by the substitution of Fe and Ni in kagome. Overall, this work serves as a guide to new design strategies for developing large AHCs and AHAs.

\section{Acknowledgments}

The authors thank V. R. Reddy and A. Gome for HRXRD,  M. Gupta and L. Behera for XRD, D. M. Phase and V. K. Ahire for EDS measurements. D.K. acknowledges the research grant from SERB India for Early Career Research Award (Grant No. ECR/2017/003350).

\onecolumngrid

\section{Supplimentary Information}

\section{Crystal growth and characterization}

\setcounter{figure}{0}
\makeatletter
\renewcommand{\thefigure}{S\@arabic\c@figure}
\makeatother

\setcounter{table}{0}
\makeatletter
\renewcommand{\thetable}{S\@arabic\c@table}
\makeatother

\begin{figure}[!htb]
\begin{centering}
\includegraphics[width=1\columnwidth]{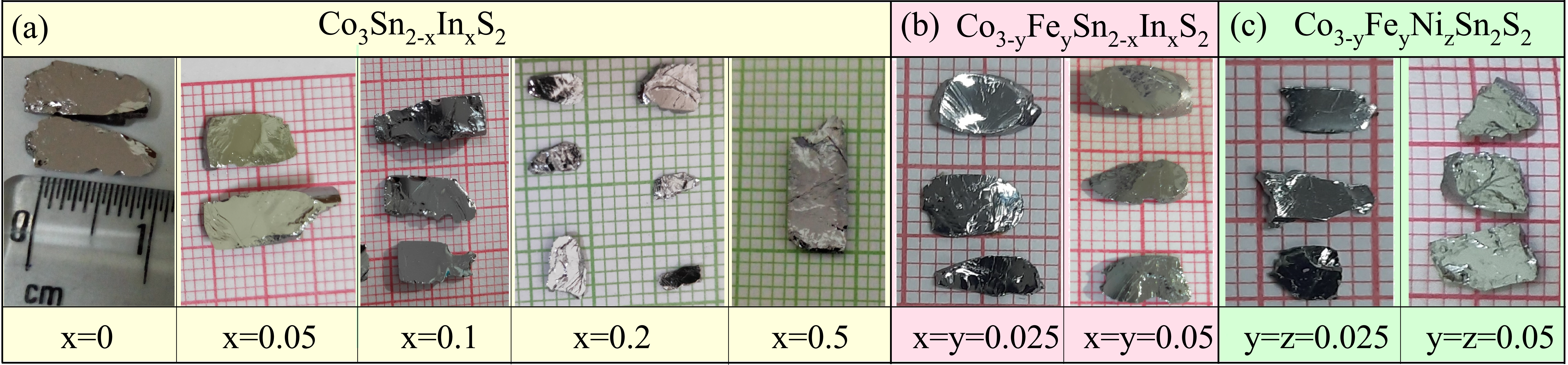}
\par\end{centering}
\caption{(Color Online) Optical image of cleaved crystals of (a) Co$_3$Sn$_{2-x}$In$_x$S$_2$ (0$\le$x$\le$0.5), (b) Co$_{3-y}$Fe$_y$Sn$_{2-x}$In$_x$S$_2$  (0$\le$x=y$\le$0.05), and (c) Co$_{3-y-z}$Fe$_y$Ni$_z$Sn$_2$S$_2$  (0$\le$y=z$\le$0.05).} \label{S1}
\end{figure}

We have grown single crystals of Co$_3$Sn$_{2-x}$In$_x$S$_2$, Co$_{3-y}$Fe$_y$Sn$_{2-x}$In$_x$S$_2$, and Co$_{3-y-z}$Fe$_y$Ni$_z$Sn$_2$S$_2$ by the self-flux method using the modified Bridgeman technique~\cite{99Rathodd2020, 99Rathod2020}. High purity Co, Sn, S, Fe, In, and Ni elements (purity\textgreater99.99\%) were taken in a stoichiometric ratio and sealed in a quartz tube under a high vacuum ($\sim$10$^{-7}$mbarr). Materials were melted at 1273K for 24 hours for reactions and then cooled to 873K at a slow rate of 2.3K/h. The obtained crystals are large ($\sim$10 mm$\times$5 mm) and have shiny metallic color, optical image of some of the cleaved crystals is shown in Fig.~\ref{S1}~(a-c).

The crystals were crushed and grinded for powder x-ray diffraction using Cu-K$\alpha$ at room temperature. The Rietveld refinement is performed on XRD raw data using space group R-3m as shown in Fig.~\ref{S2}~(a-i). The space group R-3m is a special trigonal crystal system wherein a lattice can be represented in the hexagonal setting of a rhombohedral lattice, we have used the hexagonal setting for indexing XRD data~\cite{99Rathod2020}.
\appendix
\begin{figure}[!htb]
\begin{centering}
\includegraphics[width=1\columnwidth]{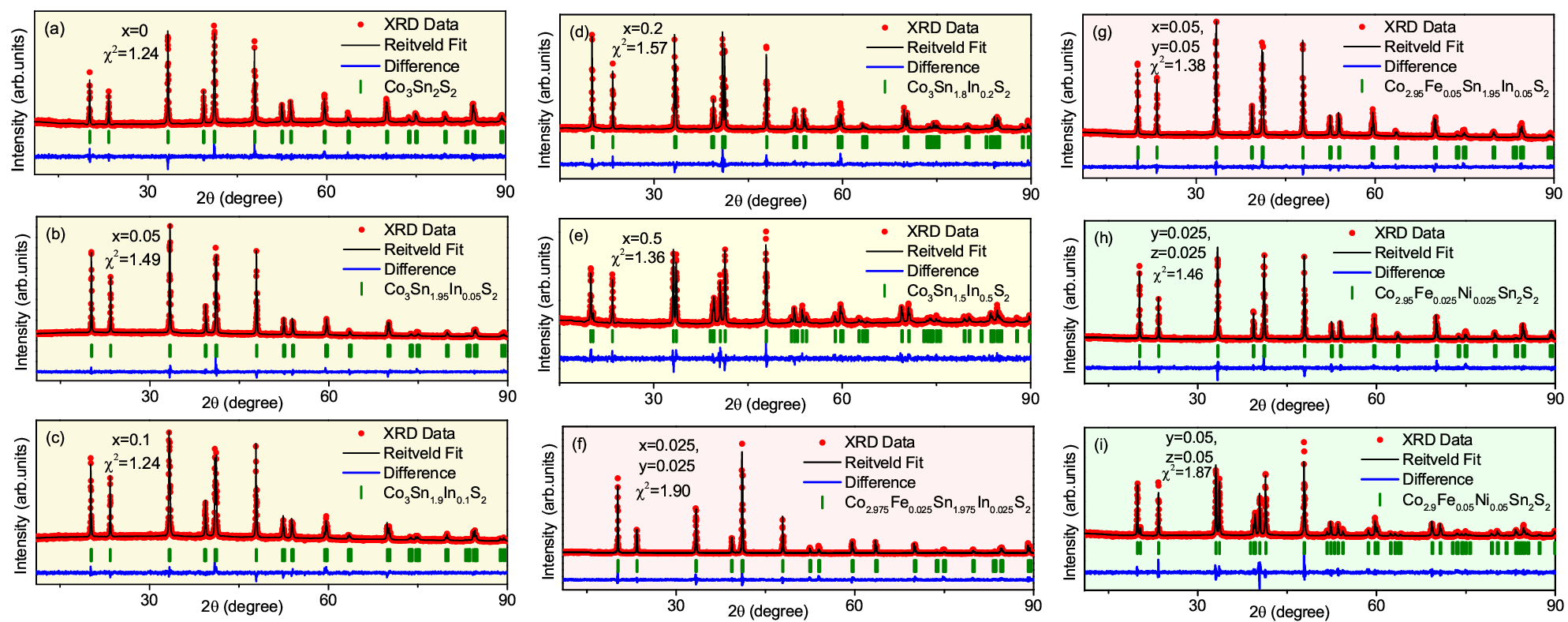}
\par\end{centering}
\caption{(Color Online) Rietveld refinement of powdered crystals of (a-e) Co$_3$Sn$_{2-x}$In$_x$S$_2$ (0$\le$x$\le$0.5), (f-g) Co$_{3-y}$Fe$_y$Sn$_{2-x}$In$_x$S$_2$  (0$\le$x=y$\le$0.05), and (h-i) Co$_{3-y-z}$Fe$_y$Ni$_z$Sn$_2$S$_2$  (0$\le$y=z$\le$0.05), where red dots corresponding to XRD raw data, the solid black line is calculated by Rietveld refinement model, blue is the difference of data and fitting, and green bars correspond to expected Bragg peaks, $\chi$$^2$ is the goodness of fit parameter.} \label{S2}
\end{figure}

\begin{figure}[!htb]
\begin{centering}
\includegraphics[width=1\columnwidth]{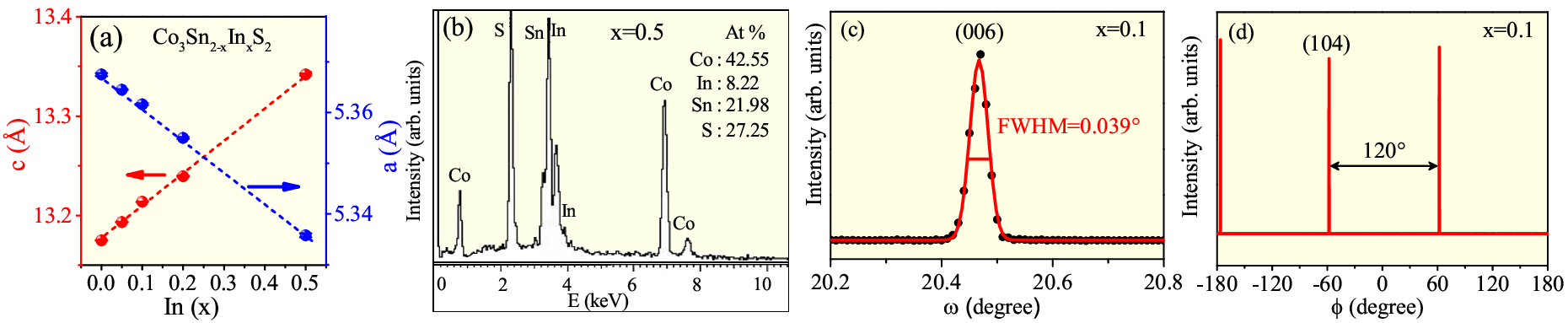}
\par\end{centering}
\caption{(Color Online) (a) Lattice parameters a and c obtained from Rietveld refinement of Co$_3$Sn$_{2-x}$In$_x$S$_2$ (0$\le$x$\le$0.5); a linear variation of parameters can be used as a reference of composition. (b) EDS spectra of  Co$_3$Sn$_{1.5}$In$_{0.5}$S$_2$ showing atomic\% of composition in good agreement with the stoichiometry of grown crystal. (c) Rocking curve or omega scan of  Co$_3$Sn$_{1.9}$In$_{0.1}$S$_2$ crystal, the fitted data showing a low full-width half maximum (FWHM) of  0.039\textdegree~corresponding to a high crystal quality~\cite{99Kumar2017}. (d) The $\phi$~ scan of (104) planes showing proper orientations along other directions and the three peaks at a separation of 120\textdegree~are related to three-fold crystal the symmetry of R-3m rhombohedral lattice~\cite{99Rathod2020}.
} \label{S3}
\end{figure}

The atomic occupancy obtained from Rietveld refinement is shown in Tab.~\ref{I}. The Rietveld refinement is quantitative and informs about the occupancy of dopants at their proper site, whereas the EDS is a qualitative technique that only informs about atomic proportions rather than the occupancy at the atomic site. In EDS, the crystals were analyzed at different regions of the same crystal to check the homogeneity, and averaged composition obtained is shown in Tab.~\ref{I}. The results are in agreement with each other indicating that dopants in grown crystals are in stoichiometric ratio, homogeneously distributed, and also occupying proper lattice sites.

\begin{table}[!htb]
\begin{tabular}{|l|l|l|}
\hline
\rowcolor{lightgray!30}\textbf{Doping content} & \textbf{Rietveld refinement occupancy} & \textbf{EDS} \\ \hline
 \rowcolor{yellow!10}x=0	& Co$_3$Sn$_{1.89}$S$_{2.04}$	& Co$_3$Sn$_2$S$_{2.09}$  \\ \hline
 \rowcolor{yellow!10} x=0.05	&  Co$_{3}$Sn$_{1.94}$In$_{0.05}$S$_2$& \\  \hline
  \rowcolor{yellow!10}x=0.1	& Co$_3$Sn$_{1.82}$In$_{0.1}$S$_{2.05}$	& \\ \hline
 \rowcolor{yellow!10}x=0.2	& Co$_3$Sn$_{1.76}$In$_{0.2}$S$_{2.04}$	& Co$_{2.94}$Sn$_{1.9}$In$_{0.26}$S$_2$\\  \hline
 \rowcolor{yellow!10} x=0.5	& Co$_3$Sn$_{1.41}$In$_{0.48}$S$_{1.89}$	& Co$_{3.08}$Sn$_{1.6}$In$_{0.58}$S$_2$\\ \hline
 \rowcolor{red!10}x=y=0.025	&Co$_{2.96}$Fe$_{0.025}$Sn$_{1.98}$In$_{0.024}$S$_{2.02}$ &\\  \hline
 \rowcolor{red!10}x=y=0.05	&Co$_{2.97}$Fe$_{0.03}$Sn$_{1.9}$In$_{0.05}$S$_{2.01}$ &\\ \hline
 \rowcolor{green!10}y=z=0.025 &Co$_{2.98}$Fe$_{0.025}$Ni$_{0.025}$Sn$_{2.01}$S$_{1.98}$ &\\  \hline
 \rowcolor{green!10} y=z=0.05	  	&Co$_{2.97}$Fe$_{0.05}$Ni$_{0.05}$Sn$_2$S$_{2.03}$ & \\ \hline
\end{tabular}
\caption{Results of Rietveld refinement occupancy for different doping concentrations of In (x), Fe (y), and Ni (z) along with averaged results of EDS  measurements for corresponding crystals.} \label{I}
\end{table}

\begin{table}[!htb]
\begin{tabular}{|l|l|l|l|l|}
\hline
\rowcolor{lightgray!30}\textbf{Doping} & \textbf{a=b(\textup {\AA})} & \textbf{c(\textup {\AA})} &  \textbf{Volume(\textup {\AA}$^3$)} & \textbf{T$_C$(K)}\\ \hline
 \rowcolor{yellow!10}x=0	         &5.3675(3)&	13.1750(3)	&328.72(2)	&      178 \\ \hline
 \rowcolor{yellow!10} x=0.05          &5.3645(2)&	13.1933(6)&328.80(2)&	172.7 \\  \hline
 \rowcolor{red!10}x=y=0.025	   &5.3662(2)&	13.1839(4)	&328.78(2)&	173.9 \\  \hline
 \rowcolor{red!10}x=y=0.05	         &5.3627(2)&	13.1936(5)	&328.59(2)&	168.4\\ \hline
 \rowcolor{green!10}y=z=0.025     &5.3666(2)&      13.1762(4)&328.64(2)&	175\\  \hline
 \rowcolor{green!10} y=z=0.05	 & 5.3278(3)&	13.3994(8)	&329.39(3)&	52.8\\ \hline
\end{tabular}
\caption{Comparison of obtained lattice parameters a, c, volume, and (T$_C$).} \label{II}
\end{table}

\begin{figure}[!htb]
\begin{centering}
\includegraphics[width=1\columnwidth]{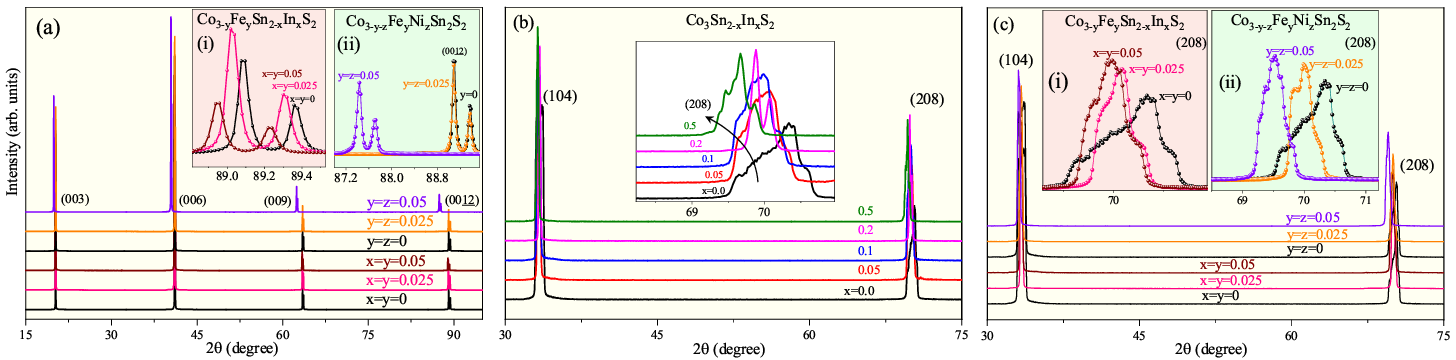}
\par\end{centering}
\caption{(Color Online) (a) Plot of out-of-plane HRXRD  $\theta$-2$\theta$ scan of single crystals of Co$_{3-y}$Fe$_y$Sn$_{2-x}$In$_x$S$_2$  and Co$_{3-y-z}$Fe$_y$Ni$_z$Sn$_2$S$_2$, indexed with [00l] planes, absence of other peak indicate high phase purity. The inset (i) and (ii) shows the magnified view of (00\underline{12}) peak. Upon doping the peak shift towards low angles due to increasing in lattice parameter c. For y=z=0.05, a relatively large change in c is observed. (b) Plot of in-plane $\theta$-2$\theta$ scan of (208) planes for Co$_3$Sn$_{2-x}$In$_x$S$_2$ demonstrating orientation of crystal domains along more than one crystallographic direction confirming single crystalline nature of the sample. The inset shows the magnified view of the (208) peak which shifts towards the lower angle with doping. Similarly for (c)  Co$_{3-y}$Fe$_y$Sn$_{2-x}$In$_x$S$_2$ and Co$_{3-y-z}$Fe$_y$Ni$_z$Sn$_2$S$_2$ as shown in inset (i) and (ii) the (208) lattice spacings systematically changing with doping.} \label{S4}
\end{figure}

\section{Resistivity and Hall data analysis}

\begin{figure}[!htb]
\begin{centering}
\includegraphics[width=1\columnwidth]{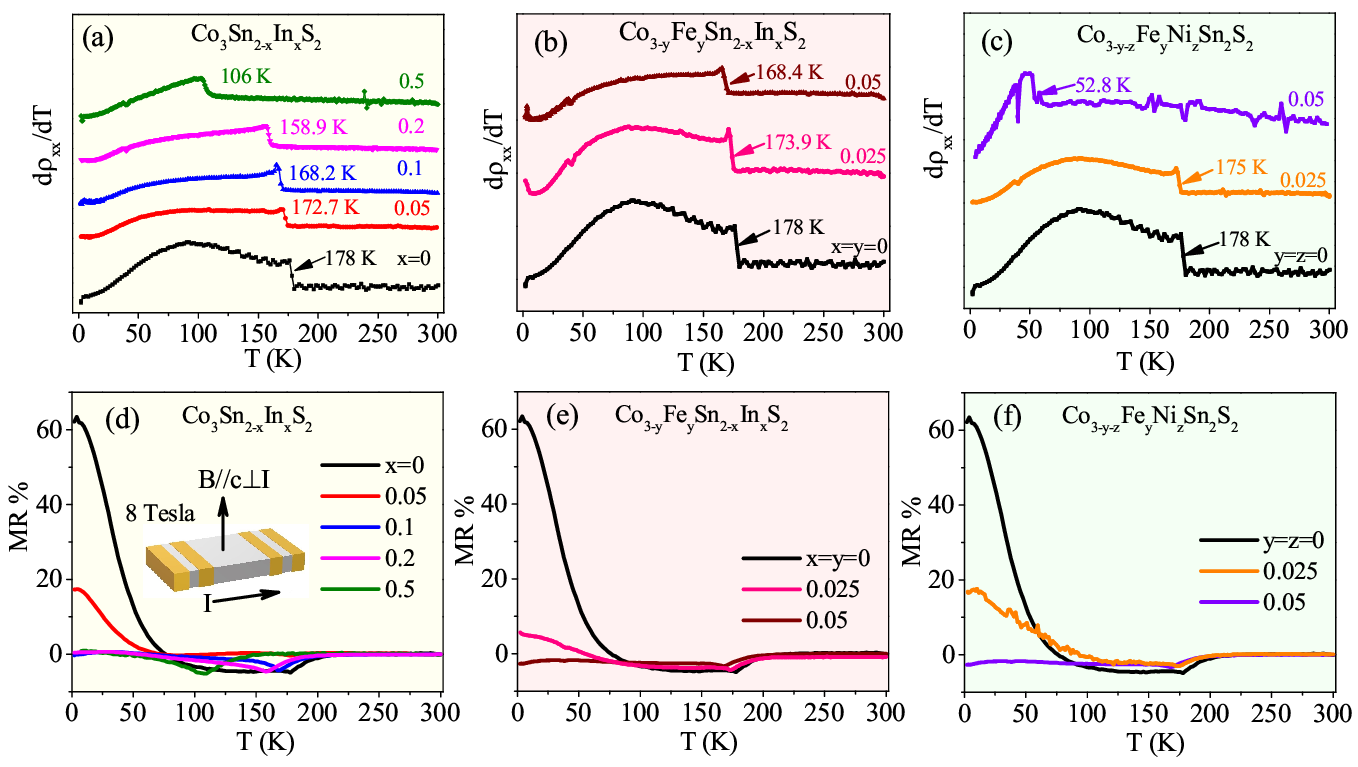}
\par\end{centering}
\caption{(Color Online) (a-c) The derivative of resistivity with temperature d$\rho$$_{xx}$/dT shows a sharp change in slope at Curie temperature (T$_C$). (d-f) Temperature dependence of magnetoresistance\%. The inset shows the schematic of the sample with the configuration of contacts; the magnetic field (B) is applied along the c-axis [003] in the direction perpendicular to current (I) (B$\parallel$c$\perp$ I).} \label{S5} \label{S6}
\end{figure}

\begin{figure}[!htb]
\begin{centering}
\includegraphics[width=1\columnwidth]{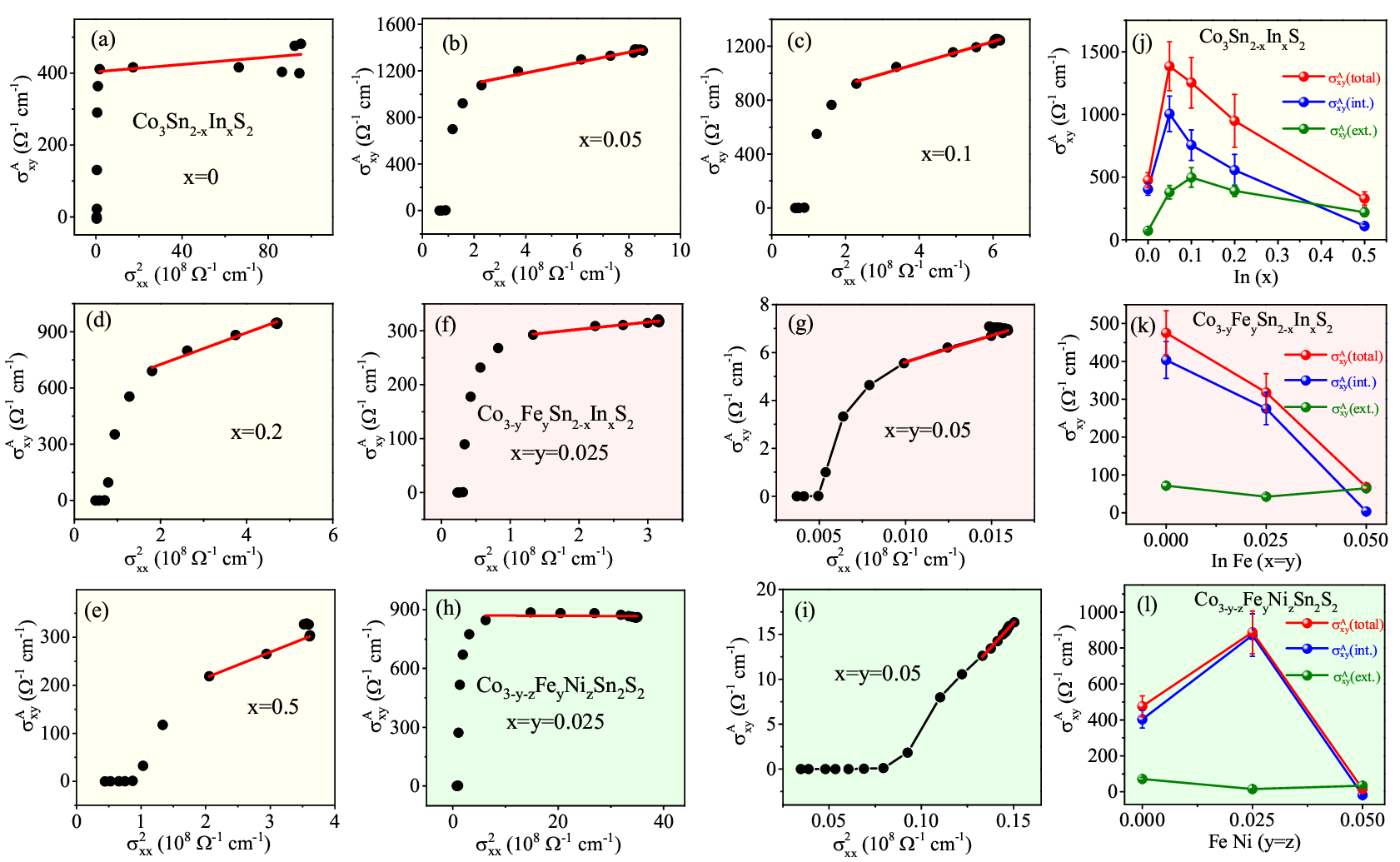}
\par\end{centering}
\caption{(Color Online) The plot of anomalous Hall conductivity ($\sigma^A_{xy}$) vs square of longitudinal conductivity ($\sigma_{xx}$) according to TYJ scaling model~\cite{99Tian2009}.  The  red lines are the linear fitting of $\sigma^A_{xy}$=-a$\sigma^{-1}_{xx0}$$\sigma^2_{xx}$-b, intercept represents intrinsic contribution, done for (a-e) Co$_3$Sn$_{2-x}$In$_x$S$_2$ from x=0 to 0.5, for (f-g) Co$_{3-y}$Fe$_y$Sn$_{2-x}$In$_x$S$_2$  at x=y=0.025 and 0.05, and for (h-i) Co$_{3-y-z}$Fe$_y$Ni$_z$Sn$_2$S$_2$ at y=z=0.025 and 0.05.  The plot of anomalous Hall conductivity ($\sigma^A_{xy}$) for (j) Co$_3$Sn$_{2-x}$In$_x$S$_2$ (k)  Co$_{3-y}$Fe$_y$Sn$_{2-x}$In$_x$S$_2$, and (l) for Co$_{3-y-z}$Fe$_y$Ni$_z$Sn$_2$S$_2$, where red balls represent total $\sigma^A_{xy}$= $\rho$$_{yx}$/[$\rho$$_{yx}$$^2$ + $\rho$$_{xx}$$^2$], the blue ball represents intrinsic anomalous Hall conductivity ($\sigma^A_{xy,in}$), and green balls represent extrinsic anomalous Hall conductivity ($\sigma^A_{xy,ex}$) obtained from TYJ analysis. The uncertainty in sample dimensions measurements gives an error in $\sigma^A_{xy}$ between (11-22\%) shown by error bars in (j-l).} \label{S7}
\end{figure}

\begin{figure}[!htb]
\begin{centering}
\includegraphics[width=1\columnwidth]{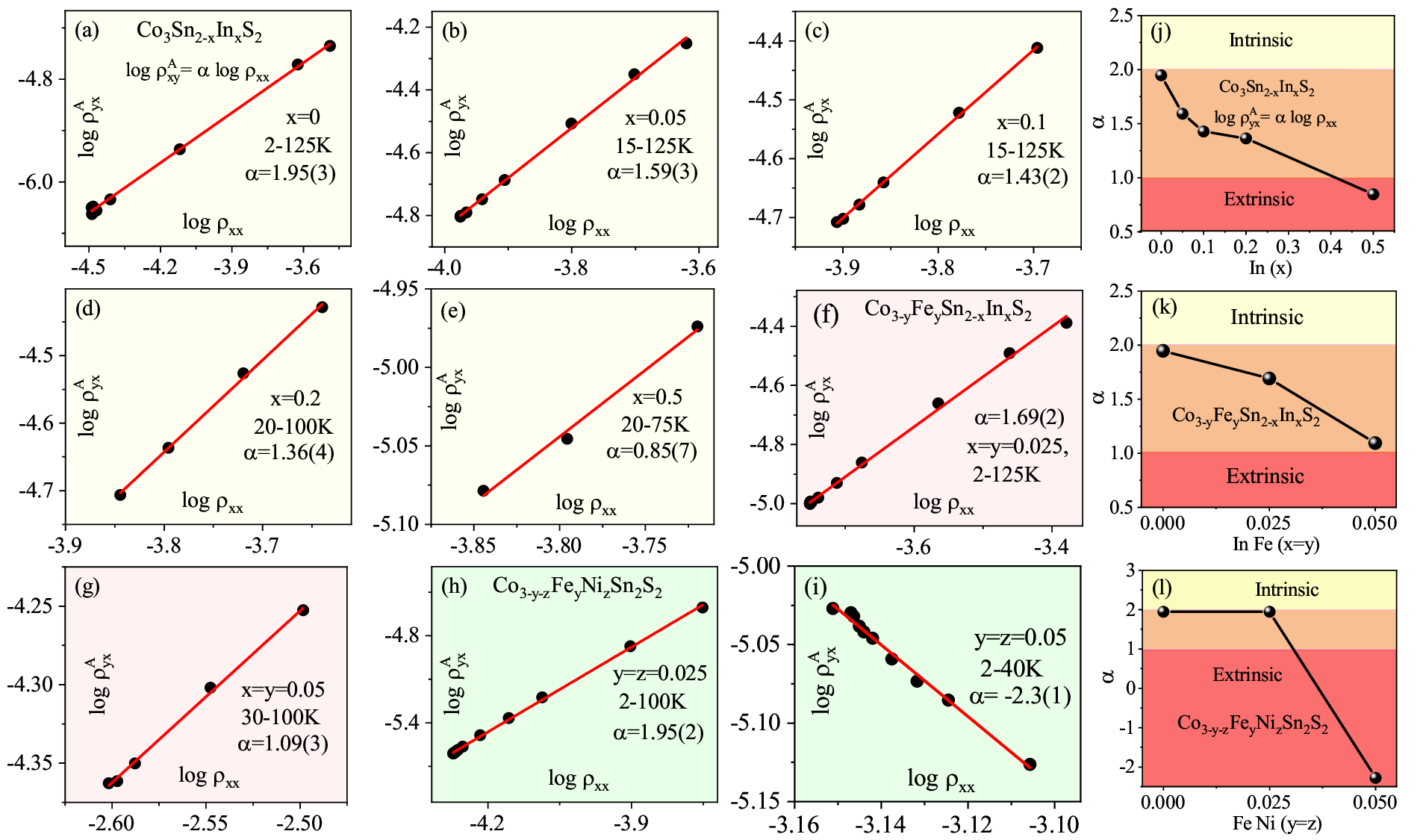}
\par\end{centering}
\caption{(Color Online)  Plot of log$\rho^A_{xy}$ vs log$\rho_{xx}$ for (a-e) Co$_3$Sn$_{2-x}$In$_x$S$_2$, (f-g)  Co$_{3-y}$Fe$_y$Sn$_{2-x}$In$_x$S$_2$, and (h-i) for Co$_{3-y-z}$Fe$_y$Ni$_z$Sn$_2$S$_2$. The red lines are the linear fit of data in the ferromagnetic region below T$_C$, and the slope of linear fit gives the value of $\alpha$ plotted in (j-l) also shown in manuscript Fig. 5(c). The yellow region ($\alpha$$\ge$2) represents the intrinsic mechanism dominated scattering behaviors of Hall resistivity, the red region ($\alpha$$\le$1) represents the extrinsic scattering mechanism dominated  Hall resistivity and the light red region (1\textless$\alpha$\textless2) is due to contribution from both intrinsic and extrinsic mechanisms.}\label{S8}
\end{figure}

\begin{figure}[!htb]
\begin{centering}
\includegraphics[width=1\columnwidth]{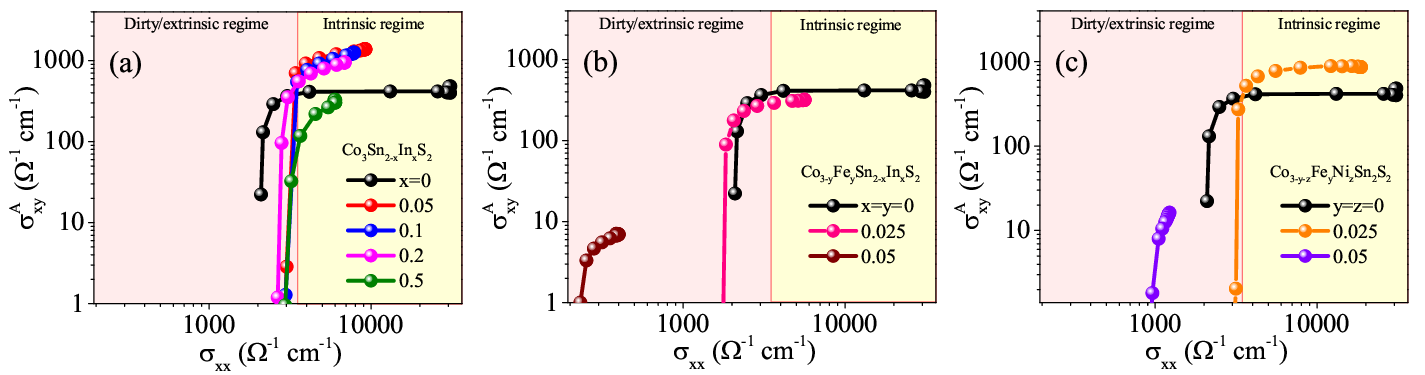}
\par\end{centering}
\caption{(Color Online) Unified model~\cite{99Onoda2006, 99Onoda2008, 99Nagaosa2010, 99Miyasato2007}  plot of anomalous Hall conductivity ($\sigma^A_{xy}$ ) vs longitudinal conductivity ($\sigma_{xx}$)  for different dopings: light red extrinsic regime (dirty) and yellow regions correspond to the intrinsic regime (moderately dirty).} \label{S9}
\end{figure}

As shown in Fig.~\ref{S9} (a-c), the pristine x=0 and y=z=0.025 samples are located toward the more conductive yellow regions to the right due to the dominance of the intrinsic mechanism. The x=y=0.05 and y=z=0.05 samples are located more toward the left light red region due to the dominance of extrinsic mechanisms, while the rest are in intermediate/boundary region signifying having contributions from both, in good agreement with TYJ analysis in Fig.~\ref{S7} and alpha values in  Fig.~\ref{S8}.

\begin{figure}[!htb]
\begin{centering}
\includegraphics[width=1\columnwidth]{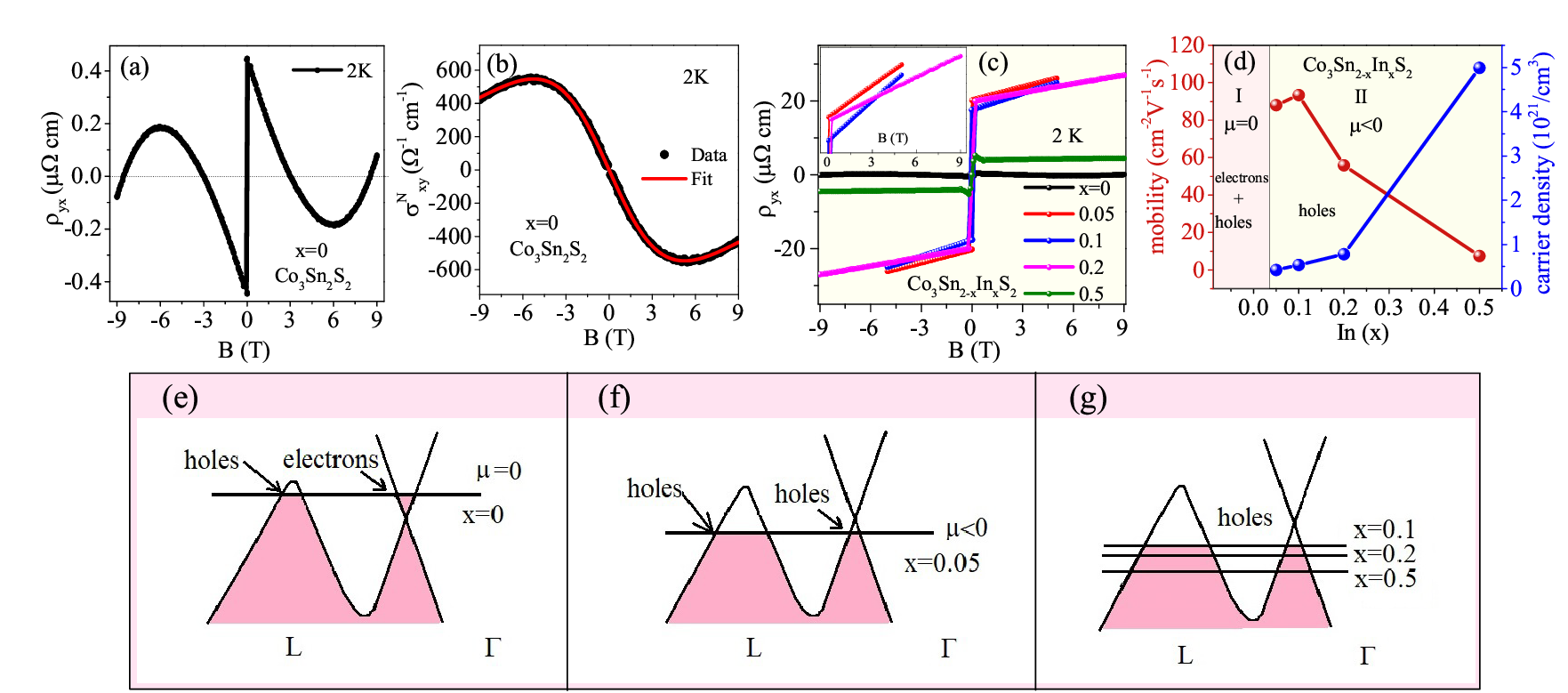}
\par\end{centering}
\caption{(Color Online) (a) The $\rho_{yx}$ at 2 K of pristine (x=0) sample shows nonlinearity. (b) The normal Hall conductivity $\sigma^N_{xy}$  fits a two-band model. (c) $\rho_{yx}$  of Co$_3$Sn$_{2-x}$In$_x$S$_2$ at 2K, the inset shows a magnified view of linearity, indicating that the electron band disappeared (d) In Co$_3$Sn$_{2-x}$In$_x$S$_2$, doping x shift the chemical potential (\textmu<0), decreases mobility~\cite{99Zhou2020} and increases carrier density. The region I, x=0 pristine (\textmu=0) where both electrons and holes contribute as shown in the band scheme (e) with increasing doping as shown in (f, g) holes dominate for a shift of chemical potential lower in energy. } \label{S10}
\end{figure}

In Fig.~\ref{S10}(a) the Hall resistivity $\rho_{yx}$ at 2K of pristine sample is a combination of holes and electrons giving positive and negative slopes, respectively. The extracted $\sigma^N_{xy}$ = $\sigma_{xy}$(B) -$\sigma^A_{xy}$(B)  is fitted using two bands Drude model~\cite{99Ziman2001}: $\sigma^N_{xy}$ =eB [n$_h$\textmu$^2_h$/(1+\textmu$^2_h$B$^2$)- n$_e$\textmu$^2_e$/(1+\textmu$^2_e$B$^2$)], as shown in Fig.~\ref{S10}(b), where \textmu$_h$ and \textmu$_e$ represent the hole and electron mobility while n$_h$ and n$_e$ represent the hole and electron carrier density. We obtained \textmu$_h$=809 cm$^2$V$^{-1}$s$^{-1}$, \textmu$_e$=1271 cm$^2$V$^{-1}$s$^{-1}$, n$_h$=1.35$\times$10$^{19}$cm$^{-3}$, n$_e$= 1.23$\times$10$^{19}$cm$^{-3}$. In pristine x=0, electron and hole carriers coexist at E$_F$, and the electron carriers have relatively higher mobility~\cite{99Zhou2020}. The band schematic of the pristine sample can be represented by Fig.~\ref{S10}(e). According to band structure calculations~\cite{99Liu2018} of Co$_3$Sn$_2$S$_2$, the E$_F$ is located near the Weyl node band crossings and gaped nodal lines. At E$_F$ a hole pocket is present at symmetry point L with electron pockets from the gaped nodal ring with band crossings~\cite{99Liu2018} between L-$\Gamma$. With hole In(x) doping as shown in Fig.~\ref{S10}(c), the slope of $\rho_{yx}$ becomes linear and positive demonstrating the dominance of holes, even for the lightest dopings. Further, as shown in Fig.~\ref{S10}(d) mobility is large around x=0.05, 0.1 and carriers are minimum and the electron bands disappear and holes dominate the transport as shown by band schematic for x=0.05 in Fig.~\ref{S10}(f).  For further higher dopings (x\textgreater0.05), shown by schematic Fig.~\ref{S10}(g), the chemical potential shifts to lower energy accompanied by an increase of carrier and decrease of mobility and MR\% shown in manuscript Fig. 3(d).

\begin{figure}[!htb]
\begin{centering}
\includegraphics[width=1\columnwidth]{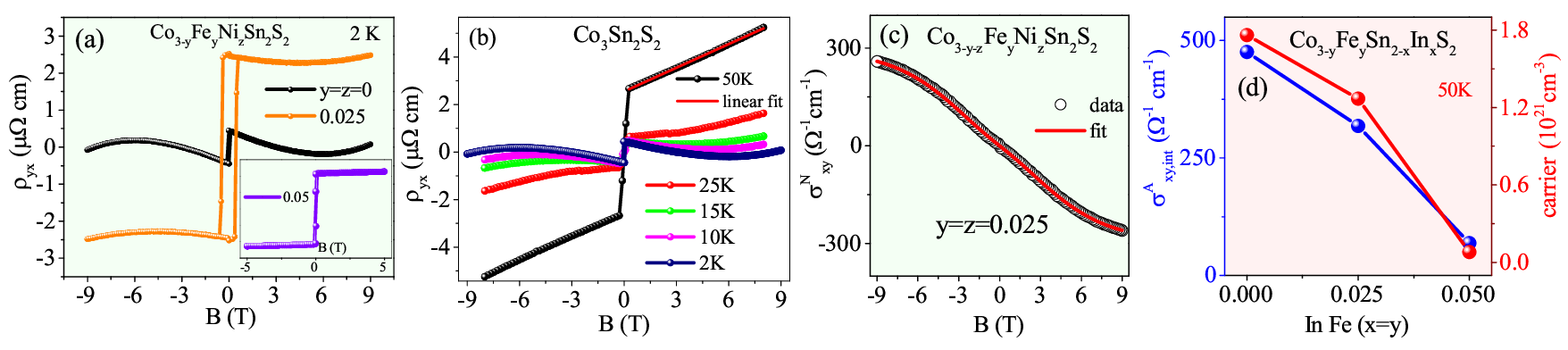}
\par\end{centering}
\caption{(Color Online) (a) $\rho_{yx}$ at 2K Co$_{3-y-z}$Fe$_y$Ni$_z$Sn$_2$S$_2$ at y=z=0 and 0.025 showing a nonlinear behavior while at higher doping y=z=0.05, $\rho_{yx}$ becomes linear as shown in the inset. (b)   Co$_3$Sn$_2$S$_2$ showing nonlinearity for T\textless50K while linear at T=50 K. (c) Two band fitting of $\sigma^N_{xy}$ at y=z=0.025. (d) Intrinsic AHC $\sigma^A_{xy,in}$, in of Co$_{3-y}$Fe$_y$Sn$_{2-x}$In$_x$S$_2$ at 50K for different dopings x=y follows hole carriers density. } \label{S11}
\end{figure}

The sample Co$_{3-y-z}$Fe$_y$Ni$_z$Sn$_2$S$_2$ is co-doped with equal electrons (Ni) and holes (Fe) which is expected to keep chemical potential fixed. As shown in Fig.~\ref{S11}(a) the $\rho_{yx}$ at y=z=0.025, maintains a nonlinearity (two-band nature) similar to the pristine sample. This nonlinearity in Co$_3$Sn$_2$S$_2$ is extremely sensitive to temperature and disappears above T\textgreater50K in the pristine sample as shown in Fig.~\ref{S11}(b) and even for light chemical doping-related chemical potential changes from x=0.05 to 0.5. The observation of nonlinearity in the co-doped y=z=0.025 sample indicates that the chemical potential has not shifted for chemical dopings. The semi-metallic and large hole \textmu$_h$=2887 cm$^2$V$^{-1}$s$^{-1}$ and electron mobilities \textmu$_e$=788 cm$^2$V$^{-1}$s$^{-1}$, and low carriers density n$_h$=9.2$\times$10$^{16}$cm$^{-3}$, n$_e$= 4.59$\times$10$^{18}$cm$^{-3}$ further suggest that occupancy of dopants and related lattice disorders have modified the Weyl bands crossing and anticrossing nodal line region present~\cite{99Liu2018} in Co$_3$Sn$_2$S$_2$ at E$_F$.

\end{document}